# Investigation of [KSF2015] 1381-19L, a WC9-type Star in the High-extinction Galactic Region

Subhajit Kar , Ramkrishna Das , and Tapas Baug
S.N. Bose National Centre for Basic Sciences, Kolkata, India; subhajit0596@gmail.com, subhajitksnbac@bose.res.in

## Abstract

We report on a multiwavelength study of the Wolf–Rayet star [KSF2015] 1381-19L, which is located in the solar metallicity region ($Z = 0.014$) of the Milky Way, strongly obscured by interstellar dust. We perform a detailed characterization of the stellar atmosphere by fitting the spectral emission lines observed in the optical and near-infrared (NIR) bands, using CMFGEN. The best-fitted spectroscopic model indicates a highly luminous ($10^{5.89} L_\odot$) star with a large radius ($15 R_\odot$) and effective temperature, wind terminal velocity, and chemical composition similar to those of Galactic WC9-dusty (WC9d)-type stars. The atmospheric ionization structure shows coexisting ionization states of different elements, simultaneously affecting the opacity and thermal electron balance. Fitting of the spectral energy data reveals high interstellar optical extinction ($A_V = 8.87$), while the IR extinction is found to be comparatively lower ($A_{K_s} = 0.98$). We do not detect any excess emission at NIR wavelengths due to dust. Upon comparison of our results with the GENEVA single-star evolutionary models ($Z = 0.014$), we identify the best possible progenitors (a rotating star of $67 M_\odot$ and a nonrotating star of $90 M_\odot$).

*Unified Astronomy Thesaurus concepts:* Wolf-Rayet stars (1806); WC stars (1793); Stellar winds (1636); Late stellar evolution (911)

*Supporting material:* data behind figure

## 1. Introduction

Wolf–Rayet (W-R) stars are a class of extremely hot ($3 \times 10^4$–$2 \times 10^5$ K) and luminous ($10^4$–$10^6 L_\odot$) Population I massive stars evolved from their O-type main-sequence progenitors ($>25 M_\odot$) via mass loss (Conti 1975) and rotation-induced mixing (Meynet & Maeder 2003). W-R stars are also formed via mass transfer through the Roche lobe overflow (RLOF) process (Paczyński 1967) or via common envelope formation, depending upon the mass of the primary companion in a massive star binary system (Crowther 2007). They exhibit strong and broad emission lines due to their dense supersonic winds. These supersonic winds deposit energy and momentum into the interstellar medium (ISM), and contribute to its chemical enrichment (Rosslowe 2016). Based upon the surface chemical composition, classical W-R stars are broadly classified into WN (He and N dominant), WC (He and C dominant), and WO (C and O dominant) types (van der Hucht 2001). These broad classes are further subclassified based on the ratios of different ionization lines of helium and nitrogen (for WN), carbon (for WC), and oxygen (for WO). WC stars not only play a significant role in the C and O enrichment of the surrounding ISM but are considered to be the progenitor of Type Ic supernova explosions (Crowther et al. 2002).

Based on the prominent C III and C IV emission lines in the optical, classical WC-spectral class objects form six subclasses: WC4–6 are WC-early (WCE) type, while WC7–9 are WC-late (WCL) type. W-R central star planetary nebulae have spectral types up to WC12. A comparative study by Crowther (2007) showed a higher population of WCL-type stars in the Milky Way (MW) than in the Magellanic Clouds due to the higher metallicity of the former than the latter. High mass-loss rates in the early stages of stellar evolution strip the outer envelopes, forming W-R stars with high wind densities, namely, WCL-type stars. These stars enrich their host galaxies through significant mass outflow. A comparative analysis showed WC-type stars in Magellanic Clouds have higher luminosity and a lower mass-loss rate than those in the MW (Crowther et al. 2002, 2023).

WC9-type stars (the coolest WCL subtype) have stronger C III features than C IV features, with a significant C II presence, and lack of O V line features in their spectra. They have similar luminosities ($10^{4.7-5.5} L_\odot$) like the WCE subtypes (as per Sander et al. 2019) but lower stellar temperature ($4 \times 10^4$–$5 \times 10^4$ K) and lower wind terminal velocity ($\sim$1300 km s$^{-1}$) than the latter. There are no conclusive studies toward developing an idea about the possible evolutionary trajectories of single classical WC9-type stars. The only possible solution for such a dilemma has been the binary evolutionary channel that predicts the typical mass of the WC object to lie between 9 and 16 $M_\odot$ (Crowther 2007). WC9-type stars, which are in a colliding-wind binary system, are incredible dust producers as they host cold and high-density regions, due to the compression of winds at the shock front, which triggers the formation of dust (Usov 1991).

Dusty WC9 binaries exist either as persistent dust sources or episodic dust sources. The former type shows no significant flux variations in the IR bands across several years, due to fixed separation between the stars in circular orbits. Also, they may sometimes depict a spiral stream of dust, for example, WR 104 with a pinwheel nebula, imaged using Keck telescope in the near-infrared (NIR) 1.65 and 2.27 μm bands (Tuthill et al. 1999). While the episodic dust producers show a periodic variation in the prominent subpeaks across the normally flat-topped He I 1.083 μm emission line during the periastron passage of the stars in an elliptical orbit, such as WR 140 (Williams et al. 1990, 2009; Varricatt et al. 2004). Such type of unique characteristics have been the core of several earlier investigations. A recent study by Williams (2019) on dusty WC9-type stars showed that most of them are of the persistent type, which tends to show large amplitude nonperiodic variability in their spectral line profiles but on short timescales (∼1−2 days). This can occur due to nonadiabatic and unstable shock-cone generated from wind–wind interaction with their binary companions (Desforges et al. 2017).

**Table 1**
Log of Spectroscopic Observations for [KSF2015] 1381-19L

| UT (hh:mm:ss) | Date | Instrument | Resolution | λ | Exposure |
|---|---|---|---|---|---|
| 18:57:33 | 2020 Jun 8 | HFOSC/Gr7 | 1400 | 3800–7000 Å | One frame ∗ 1500 s |
| 19:23:58 | 2020 Jun 8 | HFOSC/Gr8 | 2200 | 5200–9000 Å | One frame ∗ 1500 s |
| 18:30:49 | 2022 Jun 25 | TIRSPEC/$HK_s$ | 1200 | 1.50–1.84 μm | Five frames ∗ 500 s |
| ⋯ | ⋯ | ⋯ | ⋯ | 1.98–2.45 μm | ⋯ |

Recent NIR surveys (Mauerhan et al. 2009; Shara et al. 2012; Rosslowe & Crowther 2015) have revealed a larger population of W-R stars in high-extinction Galactic regions, primarily of the WCL subtype. The structure of the stellar winds and their impact on the nearby star formation region remain poorly understood. In this study, we perform a comprehensive analysis of [KSF2015] 1381-19L, a WC9-type star, located in the inner Galactic region ($6 < R_G < 9$ kpc, Rosslowe & Crowther 2015) and investigate its atmospheric stratification, which plays a crucial role in the chemodynamic evolution of the surroundings. We compare its physical properties with other WC9-type stars located in the supersolar ($Z = 0.02$) Galactic region ($R_G < 6$ kpc) and identify the prominent differences. Additionally, we search for signatures of circumstellar dust and develop a notion about its possible evolutionary channel from an O-type main-sequence star using the binary and single-star models.

### 1.1. The Target: [KSF2015] 1381-19L

Shara et al. (2009) observed the stellar objects lying in the Galactic region within $-90° < l < +60°$ and $-1° < b < +1°$ using four narrowband filters (He I 2.062, C IV 2.081, Br-γ 2.169, and He II 2.192 μm) along with the $J$ and $K_s$ bands. Follow-up observations of Kanarek et al. (2015) identified several W-R star candidates in the same Galactic region by applying an image subtraction method and confirmed their spectral types using spectroscopic observations. [KSF2015] 1381-19L was first identified as a W-R star candidate by Kanarek et al. (2015) during their spectroscopic follow-up of the Galactic region that they had first targeted during the NIR photometric surveys by Shara et al. (2009, 2012). The object is located at R.A. = $18^h12^m02\overset{s}{.}42$ and decl. = $-18°06'55\overset{''}{.}28$ and within the Galactic disk at $l = 12°.392$ and $b = 0°.167$.

Based upon the line ratios of the emission features, such as C IV 2.07–2.084 μm/C III 2.112–2.137 μm (criteria defined by Rosslowe 2016), Kanarek et al. (2015) classified the object as WC9 type as the value was estimated to be less than 1.46. Based on the GAIA DR2 data release, the object was recently added as WR 111-9 to the W-R star catalog[1] (Rosslowe & Crowther 2015) by Rate & Crowther (2020).

---
[1] http://pacrowther.staff.shef.ac.uk/WRcat/index.php

In this study, we further investigate the physical and chemical nature of the atmosphere and the surroundings of the stellar object by analyzing the observed spectroscopic data. In Section 2, we discuss the observations of the object in the optical and IR bands. The data analysis is described in Section 2.4. The spectroscopic model is discussed in Section 3 and its evolutionary phases in Section 5.2. Our concluding remarks are mentioned in Section 6.

## 2. Observations

Spectroscopic data of the object were obtained using Hanle Faint Optical spectrograph (HFOSC) in optical and TIFR InfraRed SPECtrograph (TIRSPEC; Ninan et al. 2014) in the NIR mounted on the 2 m Himalayan Chandra Telescope located at Hanle, Ladakh, India. The log of our observations is tabulated in Table 1.

### 2.1. Optical Spectroscopic Data Set

Medium resolution (∼1400–2200) optical spectra were observed with the HFOSC using grism 7 (3800–7000 Å, hereafter blue spectrum) and grism 8 (5200–9000 Å, hereafter red spectrum). A 2k × 4k CCD was used for all of the observations. FeAr and FeNe spectra were observed immediately after capturing any object spectrum for wavelength calibration. On the same night, we observed Feige 66 (Massey & Gronwall 1990), a spectrophotometric standard star through grisms 7 and 8. Details of the observation are provided in Table 1.

The optical spectra were reduced using a standard pipeline based on the IRAF (Tody 1993) package. First, we subtracted the master bias frame from each science frame and removed the cosmic rays from the bias-subtracted frames. Then, we extracted the 1D spectral data from the 2D image using the apall task in IRAF. Wavelength calibration of the object spectra was done using the dispersion-corrected lamp spectra. Flux calibration was done using the atmospheric extinction function and detector sensitivity function derived from the spectrophotometric standard star data.

### 2.2. NIR Spectroscopic Data Set

Medium-resolution ($R \sim 1200$) NIR spectra in the $H$ (1.5–1.84 μm) and $K_s$ bands (1.98–2.45 μm) were observed using the TIRSPEC instrument mounted on the Himalayan Chandra Telescope (Ninan et al. 2014). The observations were conducted in the cross-dispersed configuration, allowing us to cover two NIR wave bands simultaneously. Slit-spectroscopic observations were conducted utilizing an ABBA slit-nodding pattern at two dithered positions.

We captured five consecutive frames, each with a 500 s exposure to achieve a substantial signal-to-noise ratio

**Table 2**
Photometric Data Set of [KSF2015] 1381-19L at Different Wave Bands

| Band | Apparent Magnitude (mag) | $\lambda_c$ (nm) | Flux Density (Jy) | Database |
|---|---|---|---|---|
| (1) | (2) | (3) | (4) | (5) |
| B | 15.48 | 444 | $7.50e^{-4}$ | NOMAD |
| V | 13.86 | 554 | $2.45e^{-3}$ | NOMAD |
| G | 13.60 | 582 | $1.16e^{-2}$ | GAIA |
| J | 9.66 | 1240 | $2.15e^{-1}$ | 2MASS |
| H | 8.62 | 1650 | $3.74e^{-1}$ | 2MASS |
| $K_s$ | 7.80 | 2160 | $5.14e^{-1}$ | 2MASS |
| W1 | 7.34 | 3350 | $3.55e^{-1}$ | WISE |
| IRAC — 3.6 | 7.17 | 3550 | $3.8e^{-1}$ | SPITZER |
| W2 | 6.62 | 4600 | $3.83e^{-1}$ | WISE |
| IRAC — 5.8 | 6.34 | 5730 | $3.32e^{-1}$ | SPITZER |
| IRAC — 8.0 | 6.01 | 7870 | $2.53e^{-1}$ | SPITZER |
| W3 | 5.90 | 11,600 | $1.26e^{-1}$ | WISE |
| W4 | 4.89 | 22,100 | $9.2e^{-2}$ | WISE |

(S/N) ∼ 40. Argon calibration lamp frames and tungsten continuum flats were observed immediately after observing each data set. A relatively bright and nearby located telluric standard star HD 172792 (H mag = 8.77, A0V type) was observed using the same slit across all the spectral orders for telluric correction. Also, dark frames were observed before and after the observation on the same night. The log of the observations is given in Table 1.

For data extraction, we used the publicly distributed semi-automated TIRSPEC pipeline (Ninan et al. 2014). Object frames were median combined to remove the cosmic-ray outliers and to enhance the S/N. To generate flat corrected object frames for both dither positions, the median combined object frame was divided by the corresponding normalized median combined continuum flat frame. Further, the flat and cosmic-ray corrected object frames of both dithered positions (A and B) were subtracted from each other (in A-B and B-A format) to remove the sky background. Finally, a 1D spectral image of the object was extracted and wavelength calibrated with the corresponding Argon lamp spectra.

The final spectra were generated after *average* combining the extracted and wavelength-calibrated spectrum of each dithered pair. The telluric-corrected science spectrum was flux calibrated using the Two Micron All-Sky Survey (2MASS) photometric magnitudes (Table 2) of the H and $K_s$ bands of the science object.

### 2.3. Photometric Data Set

We retrieved optical and IR photometric flux data and apparent magnitudes from the online accessible VizieR data catalog,[2] which serves as a repository for diverse databases. The optical B- and V-band flux data utilized in the analysis are retrieved from the I/297 catalog, which hosts the Naval Observatory Merged Astrometric Dataset (NOMAD; Zacharias et al. 2004). The optical G-band data are acquired from the I/355/gaiadr3 catalog (Gaia Collaboration 2022), hosting GAIA DR3 (Gaia Collaboration et al. 2023) data. The NIR data are retrieved from the II/246 catalog (Cutri et al. 2003), hosting the 2MASS database (Skrutskie et al. 2006). Additionally, mid-infrared (MIR) data are sourced from the II/328/allwise catalog (Cutri et al. 2021) hosting Wide-field Infrared Survey Explorer (WISE) data (Wright et al. 2010) and the II/293/glimpse catalog (Spitzer Science Center 2009), hosting Spitzer-IRAC band data from the GLIMPSE-I survey (GLIMPSE Team 2020). The integrated flux density values observed in each photometric band, ranging from the optical to the MIR, are presented (with the respective databases) in Table 2.

### 2.4. Data Analysis

To classify and characterize the spectroscopic nature of the object, we derive the spectral parameters, such as the equivalent width (EW) and full width at half-maximum (FWHM), associated with every emission line present in the spectra. Before measuring the EW and FWHM, we define the spectral continuum for each spectral band. To identify the continuum, we choose regions that lack spectral features. We use the conventional polynomial fitting procedure using the Levenberg–Marquardt (Levenberg 1944; Marquardt 1963, hereafter LM) algorithm and divide the flux-calibrated spectrum with the best-fitted continuum. Similarly, we apply this method for normalizing the spectra in the NIR bands, which are affected by excess emissions due to free–free emissions and sometimes circumstellar dust.

We estimate the values of the EW and FWHM associated with the emission lines, using the astronomical Python libraries, which included several `Astropy`[3] (Astropy Collaboration et al. 2013, 2018, 2022) modules such as `models`, `units`, `io`, and `specutils` modules, such as `spectra`, `fitting`, `analysis`, `SpectralRegion`, and `manipulation`. For our analysis, we extract the spectral region (using the `extract_region` task of the `manipulation` module) containing any emission line(s) and fit the line profiles of each emission line using the 1D Gaussian function (using the `fit_lines` task of the `fitting` module) and determine the best profile fitting using the LM chi-square minimization technique. We estimate the EW of the spectral lines from the profile-fitted model.

We calculate the normalized mean flux for both the spectral line ($\overline{F_{\rm line}}$) and the continuum ($\overline{F_{\rm cont}}$). The continuum of the spectrum is chosen from a featureless spectral region. We also estimate the S/N of the spectra from the considered pseudo-continuum region. We estimate the error in the EW of the spectral line using the technique derived in Section 3.2 of Vollmann & Eversberg (2006).

Following the method discussed in Section 3.1 of Zhekov et al. (2014), we remove the broadening effect of the spectrograph from the measured FWHM values using the spectral resolution of the corresponding dispersing elements in case of optical as well as NIR wave bands (Table 1). The parameter values corresponding to each identified emission line are presented in Tables 3 and 4.

#### 2.4.1. Optical Spectrum

In the bluer part of the spectrum (3800–5000 Å), we identify only a few emission lines with low S/N due to the faintness of the target object (with B-mag ∼ 16). We could only detect C III $\lambda$4651 and He II $\lambda$4686 emission lines within the 4000–5000 Å of the blue spectrum, as can be seen in Figure 1(a). However, no other physical diagnostic emission lines (van der Hucht 2001) such as C II $\lambda$4267, C III $\lambda$4326, He II $\lambda$4339,

---

[2] https://vizier.cds.unistra.fr/

[3] https://www.astropy.org/

**Table 3**
Observed EW and FWHM of Emission Lines in Optical Spectrum

| Wavelength (Å) | Atomic Species | FWHM (Å) | EW (Å) | Transition |
|---|---|---|---|---|
| (1) | (2) | (3) | (4) | (5) |
| 4651 | C III | 19.70 | 109 ± 15 | $3p\ ^3P_0^o - 3s\ ^3S_1^e$ |
| 4686 | He II | 17.99 | 36 ± 6 | 4–3 |
| 5696 | C III | 33.66 | 454 ± 22 | $3d\ ^1D_2^e - 3p\ ^1P_1^o$ |
| 5802/12 | C IV | 29.08 | 124 ± 7 | $3p\ ^2P_{3/2}^o - 3s\ ^2S_{1/2}^e$ |
| 5876 | He I | 24.82 | 78 ± 5 | $3d\ ^3D^e - 2p\ ^3P^o$ |
| 6578 | C II | 45.26 | 192 ± 10 | $3p\ ^2P_{3/2}^o - 3s\ ^2S_{1/2}^e$ |
| 6681 | He I | 38.48 | 43 ± 4 | $3d\ ^1D^e - 2p\ ^1P^o$ |
| 6727–73 | C III | 31.08 | 125 ± 10 | $3p\ ^3D^e - 3s\ ^3P^o$ |
| 7063/68 | C IV | 30.32 | 73 ± 5 | $9z\ ^2Z - 7z\ ^2Z$ |
| 7231 | C II | 42.55 | 267 ± 13 | $3d\ ^2D_{3/2}^e - 3p\ ^2P_{1/2}^o$ |
| 7487 | C III | 66.13 | 32 ± 6 | $5d\ ^3D^e - 6f\ ^3F^o$ |
| 7727 | C IV | 38.35 | 31 ± 4 | $11z\ ^2Z - 8d\ ^2D^e$ |
| 8197 | C III | 29.41 | 56 ± 5 | $6h\ ^3H^o - 5g\ ^3G^e$ |
| 8348 | C III | 59.07 | 125 ± 12 | $3d\ ^3F_3^o - 4d\ ^3D_2^e$ |
| 8500 | C III | 39.49 | 70 ± 5 | $3p\ ^1P_1^o - 3s\ ^1S_0^e$ |
| 8664 | C III | 37.80 | 54 ± 5 | $6w\ ^2W - 6z\ ^2Z$ |

O II λ4416, C IV λ4441, He I λ4472, and C III λ4517 are detected in the blue spectrum.

We find the emission lines in the red spectrum (5200–9000 Å) to be prominent with better S/N. Consistent with the characteristics of a WC9-type star, we observe robust and broad emission line features primarily originating from the lower ionization states of carbon and helium. We identify the prominent optical emission lines that are formed in the inner layers of the expanding atmosphere, such as C III λ5696, λλ6727–6773, λ7487, 8348, 8500, 8664, C IV λλ5802–12, 7063-68, He II λ5411, and O III λ5592. Furthermore, we observe emission lines that originate at less ionized atmospheric regions (Hillier 1989), such as He I λ5876, 6681 and C II λ6578, 7231. These lines provide important information regarding the structure and dynamics of the outer layers of the expanding atmosphere. The C III λλ6727–6773 line is blended with the adjacent He I λ6768 line.

The spectral parameters for each line are estimated using the method discussed in Section 2.4. For our analysis, we select the pseudo-continuum region from 4700–4850 Å for the blue spectrum and 5920−6120 Å for the red spectrum, as these regions are free from significant spectral features. The S/N is computed from the chosen pseudo-continuum regions in the case of the blue spectrum (∼20) and the red spectrum (∼28). In Table 3, we present the spectral parameters and the corresponding statistical errors of the line EW for each observed emission line in the optical spectra. Additionally, the atomic transitions of the emission lines are presented in column 5 of Table 3.

#### 2.4.2. NIR Spectrum

The NIR spectral region of the W-R stars consists of crucial emission lines originating due to higher level atomic transitions at a larger stellar radius than the optical and UV radiative zones (Rosslowe 2016). Therefore, the line strengths and profiles may provide significant information about the outer wind geometry and the effect of clumping on the mass-loss rates.

From the observed spectra, we identify prominent emission line features of He II 1.572 μm, He I 1.701 μm, and C IV 1.736 μm in the $H$ band. In the $K_s$ band, we detect He I 2.059 μm, C IV 2.070–2.084 μm triplet, C III+He I 2.112–2.137 μm, He II 2.165 μm, He II 2.189 μm, and C III 2.325 μm emission lines. We estimate spectral parameters (EW and FWHM) of the corresponding emission lines. For our analysis, we select the pseudo-continuum region from 1.520–1.543 μm in the $H$ band and 2.212−2.232 μm in the $K_s$ band as they are free from significant spectral features. The S/N (∼35) is computed from these continuum regions. The spectral parameters and their corresponding statistical errors for each emission line in the $H$ and $K_s$ bands are presented in Table 4. Also the measured line strengths of [KSF2015] 1381-19L in the $K_s$ band from Kanarek et al. (2015) are given in Table 4.

### 3. Spectroscopic Modeling Procedure

To understand the nature of the stellar atmosphere, we perform quantitative analysis of the observed optical and NIR spectra using suitable modeling techniques.

#### 3.1. Methodologies

For data modeling, we have used a 1D-radiative transfer modeling code CMFGEN (Hillier & Miller 1998; Hillier 2003, 2012), which solves the radiative transfer equations for a spherically expanding atmosphere in the comoving frame subject to the statistical and radiative equilibrium conditions. The modeling code applies a linearization technique to achieve consistency among the temperature structure, radiation field, and atomic level populations. A converged model is generated iteratively as the radiation field and level populations are dependent on each other explicitly.

As the W-R stars have an optically thick atmosphere accelerated by the UV radiation pressure, it is difficult to accurately determine the inner radius of the stellar atmosphere, due to increasing optical depth. For a general agreement with the stellar evolutionary works, the radius of the star ($R_*$) is considered to be located at a Rosseland optical depth of 20, and its effective temperature ($T_*$) is estimated from the Stefan–Boltzmann's law. In the case of W-R stars, the temperature and radius of the photosphere are considered at an optical depth of two-thirds (from gray approximation).

Multiple studies (Hillier 1989; Sander et al. 2012; Aadland et al. 2022) showed that for a WC-type star, the stellar wind velocity in the case of broad-lined stars follows a two-component beta velocity law (Hillier 2003):

$$v(r) = \frac{v_{\text{phot}} + (v_{\infty,2} - v_{\text{phot}})\left(1 - \frac{R_*}{r}\right)^{\beta_1} + (v_{\infty,2} - v_{\infty,1})\left(1 - \frac{R_*}{r}\right)^{\beta_2}}{1 + (v_{\text{phot}}/v_{\text{core}} - 1)\exp[(R_* - r)/\Delta R]},$$

(1)

where $v_{\text{phot}}$ denotes the photospheric velocity; $v_{\text{core}}$ denotes the velocity at the stellar core ($R_*$ at $\tau_{\text{ross}} = 20$); $v_\infty$ denotes the terminal velocity that is considered to be the maximum velocity the stellar wind can achieve at the largest radii; $\Delta R$ denotes the scale height of photosphere; and $\beta$ denotes the acceleration exponent that defines how fast the wind is expanding. As the higher excitation lines exhibit slightly narrower profiles than the lower excitation lines, a small value for $\beta_1$ (e.g., 0.8, 1.0, etc., Hillier 1991; Sander et al. 2012) is selected in the inner

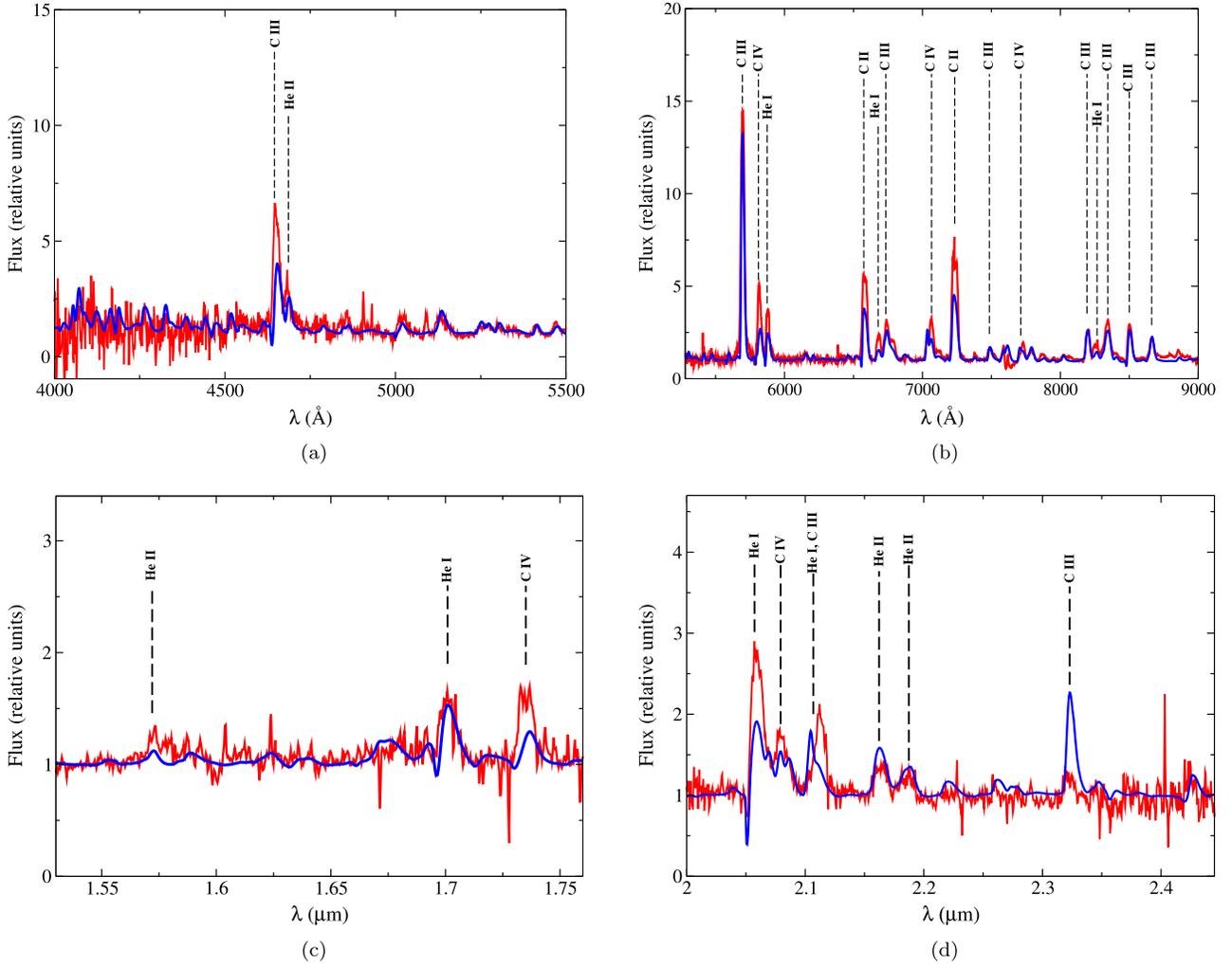

**Figure 1.** Fitting of model spectra (represented in blue) with the observed (represented in red) across optical in Gr7 (a) and Gr8 (b); NIR $H$ (b) and $K_s$ (c) bands. The associated data is available for download in the online journal.

(The data used to create this figure are available.)

Table 4
Observed EW and FWHM of Emission Lines in the NIR ($H$ and $K_s$) Bands

| NIR Band | Wavelength | Atomic Species | This Study | | KSF (2015) Study | | Transition |
|---|---|---|---|---|---|---|---|
| | | | FWHM | EW | FWHM | EW | |
| | ($\mu$m) | | (Å) | (Å) | (Å) | (Å) | |
| (1) | (2) | (3) | (4) | (5) | (6) | (7) | (8) |
| | 1.572 | He II | 41 | 15 ± 3 | ⋯ | ⋯ | 13–7 |
| $H$ | 1.701 | He I | 71.74 | 49 ± 7 | ⋯ | ⋯ | $4d\ ^3D^e$–$3p\ ^3P^o$ |
| | 1.736 | C IV | 62.25 | 48 ± 6 | ⋯ | ⋯ | $9z2Z$–$8z2Z$ |
| | 2.059 | He I | 85.36 | 99 ± 12 | 63.3 | 88.10 | $2p\ ^1P^o$–$2s\ ^1S^e$ |
| | 2.078 | C IV | 169.32 | 181 ± 14 | 193.5 | 170.15 | $3d\ ^2D^e_{5/2}$–$3p\ ^2P^o_{3/2}$ |
| $K_s$ | 2.117 | C III | 103.09 | 110 ± 11 | 224.5 | 123.28 | $8h\ ^3H^o$–$7i\ ^3I^e$ |
| | 2.165 | He II | 100.30 | 44 ± 8 | 117.9 | 40.83 | 14–8 |
| | 2.189 | He II | 111.20 | 32 ± 8 | 160.5 | 43.35 | 10–7 |
| | 2.217 | He II | ⋯ | ⋯ | 156.4 | 61.83 | 22–9 |
| | 2.325 | C III | 76.38 | 20 ± 5 | ⋯ | ⋯ | $5p\ ^3P^o$–$5s\ ^3S^e$ |

winds. Conversely, a larger value for $\beta_2$ (e.g., 20, 50, etc., Hillier 2003) is chosen for the outer winds, where there is lesser wind acceleration. The presence of redshifted wings is attributed to electron scattering. They preferentially occur on the red side of the profile as the wind velocities exceed the thermal velocities of the electrons (Aadland et al. 2022).

Multiple studies (Owocki et al. 1988; Hillier 1991; Crowther et al. 2006; Williams et al. 2015) have confirmed the presence of clumpiness in the stellar winds of W-R stars. CMFGEN incorporates clumping using velocity-dependent volume filling factor through the following relation:

$$f = f_\infty + (1 - f_\infty)\exp\left(\frac{-v(r)}{v_\text{clump}}\right), \quad (2)$$

where $f_\infty$ signifies the filling factor at a larger radius, and $v_\text{clump}$ is the rate at which clumping varies. The clumping parameters are chosen to simplify the size of the clumps at different radii in the winds (mainly in the outer regions). CMFGEN models are computed upon considering the medium between the clumps as vacant. Although small in size, the effects of clumps were seen on the electron scattering wings (Hillier 1991) and varying subpeaks on the spectral line profiles (Moffat et al. 1988; Lépine & Moffat 1999; Lépine et al. 2000). However, it has been noted earlier (Hamann & Koesterke 1998) that the line strength of the emission lines remains unchanged when the mass-loss rate is reduced by a factor of $\sqrt{f}$. Clumps affect the wind acceleration process as denser clumps can hinder the acceleration (Schmutz 1997) of the stellar wind, leading to a lower terminal velocity.

In WC stars, the emission lines and continuum are generated due to the combined effects of several mechanisms, which are simultaneously addressed by CMFGEN, such as excitation due to thermal collisions (e.g., C IV $\lambda$1549, C III $\lambda$1909), radiation-based recombination (e.g., C IV $\lambda$5471, He II $\lambda$5411), effects of fluorescence on the continuum (e.g., C IV $\lambda\lambda$5802–12), and recombination at low-temperature due to dielectronic transitions (e.g., C III $\lambda\lambda$2297, 4649, 5696, 6741, 9710) (Hillier 1989).

The atomic data containing the line and continuum opacities of various ionization states of several elements are chosen to obtain sufficient radiative driving of the stellar winds. To account for the wind blanketing effect, the super-level approach (Gräfener et al. 2002) is considered for the dominant elements: He, C, O, and Fe as they undergo complex atomic transitions in the winds. The atomic data source for the energy levels, photoionization cross sections, oscillator strengths, collision strengths, dielectronic recombination data, and autoionization states are listed in Table 5.

## 4. Results

Spectral line profile fitting is an interplay between several physical parameters, such as luminosity, temperature, mass-loss rate, and surface chemical composition. The stellar luminosity controls the wind ionization structure and simultaneously affects the EWs of the emission lines of different ionization states of the same atomic species. The mass-loss rate affects the wind density ($\rho$) dependent recombination lines. The stellar winds get heated by the photoionization mechanism. In Figure 2(b), it can be seen that the temperature decreases rapidly in the inner wind, due to effective cooling from the radiative recombination processes. The temperature of the outer wind falls almost exponentially with decreasing electron density due to the cooling mechanism by the free–free and thermal collisional transitions (Hillier 1989). The line strengths of the higher ionization emission lines indicate the stellar temperature, whereas the line widths of the lower ionization

Table 5
Source of Model Atomic Data

| Type (1) | Database (2) |
| --- | --- |
| Energy levels | NIST (Kramida et al. 2023) |
| | Kurucz (2009, 2010) |
| Photoionization cross section | OPACITY Project (Seaton 1987) |
| | NORAD (Nahar 1998, 2010) |
| | Kurucz (2009, 2010) |
| | Davey et al. (2000) |
| | PJ Storey (2012) |
| | Luo & Pradhan (1989) |
| | Iron Project data (Hummer et al. 1993) |
| Oscillator strengths | OPACITY Project (Seaton 1987) |
| | Leibowitz (1972) |
| | Luo & Pradhan (1989) |
| | Kurucz (2009, 2010) |
| Collision strengths | Cochrane & McWhirter (1983) |
| | Tayal (2008) |
| | Mendoza (1983) |
| | Chen & Pradhan (1999) |
| | Zhang et al. (1994) |
| | Zhang & Pradhan (1997) |
| Dielectronic data | Nussbaumer & Storey (1983) |
| Autoionization data | Safronova et al. (1998) |

states provide an idea about the terminal velocity and the turbulent velocity of the wind.

Spectral lines in W-R stars are blends of several adjacent lines across different elemental species. We ensure that all types of atomic transitions, including the low- and high-temperature dielectronic recombinations from doubly excited autoionized states (Hillier 2011) are also covered. We compare models with different sets of atomic data to determine the best set of atomic levels that generate most of the spectroscopic features in the model. For our model, we consider a suitable set of atomic data similar to Rosslowe (2016) containing the following atomic species: He I–II, C II–IV, O II–IV, Ne II–IV, Si III–IV, S III–VI, Ar III–V, Ca II–VI, and Fe III–VI. The number of atomic levels for each ionized species used in our model is constrained from the best-fitted spectroscopic model. Our atomic model consists of 985 super levels and 3052 full levels, which undergo about 45,000 non-LTE transitions. The list of atomic levels adopted for each atomic ionization state of the model atomic species is tabulated in Table 6.

The stratified winds (Figure 3(a)) show that helium remains doubly ionized ($He^{+2}$) until $N_e > 10^{11}\,\text{cm}^{-3}$, while the population of neutral $He^0$ increases outward ($>100\,R_*$), leading to stronger line strengths in the IR bands. Weak strengths and a fewer number of C IV transitions are due to the confined formation zone of $C^{+3}$ ions (see Figure 3(b)). The relative population of $C^{+2}$ is enhanced from $N_e \sim 3 \times 10^{11}\,\text{cm}^{-3}$, due to recombination transitions from C IV–C III. At the outer wind, $C^{+2}$ becomes dominant (as $N_e > 10^9\,\text{cm}^{-3}$), indicating a broad extent of C III lines. The $O^{+2}$ ions dominate within $\tau_\text{ross} = 0.1$–0.01, due to strong O IV–O III recombinations. At lower optical depth ($\tau_\text{ross} > 0.01$), $O^{+2}$ radiatively decays to $O^{+1}$ ions.

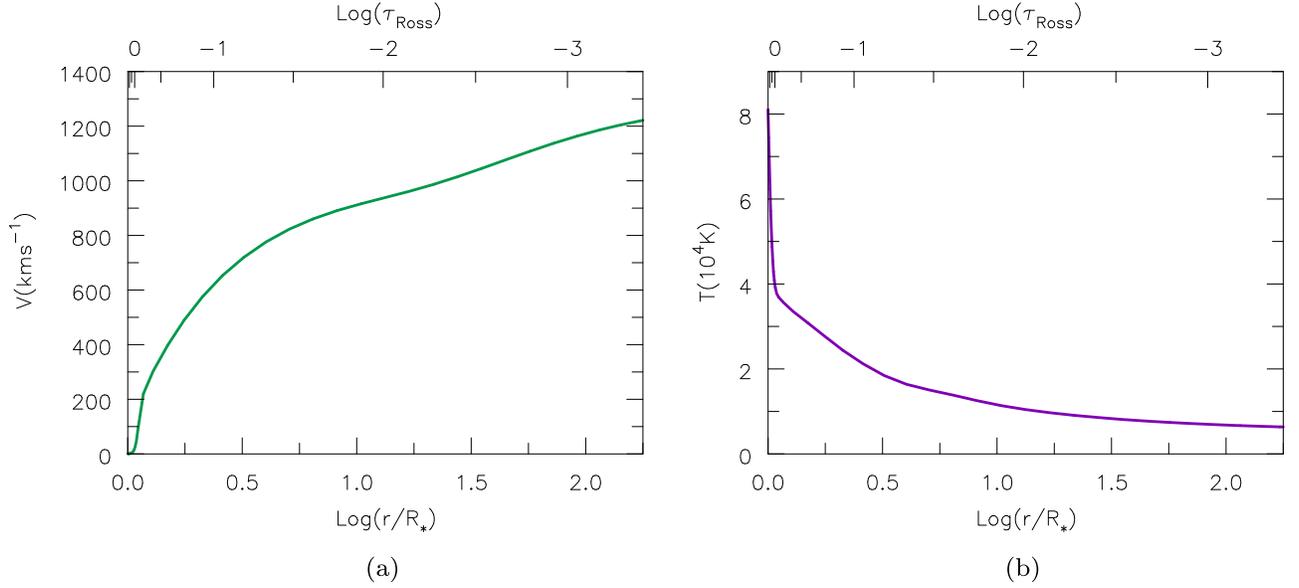

**Figure 2.** (a) The wind velocity and (b) temperature profile of the stellar atmosphere (see Section 4.1).

Table 6
Atomic Models Used for the Analyses

| Atomic Species | $N_S$ | $N_F$ |
|---|---|---|
| He I | 45 | 69 |
| He II | 22 | 30 |
| C II | 104 | 338 |
| C III | 81 | 141 |
| C IV | 59 | 64 |
| O II | 46 | 102 |
| O III | 44 | 80 |
| O IV | 41 | 60 |
| Ne II | 22 | 96 |
| Ne III | 31 | 102 |
| Ne IV | 17 | 52 |
| Si III | 25 | 45 |
| Si IV | 22 | 33 |
| S III | 25 | 41 |
| S IV | 36 | 77 |
| S V | 29 | 31 |
| S VI | 18 | 20 |
| Ar III | 25 | 150 |
| Ar IV | 31 | 107 |
| Ar V | 25 | 50 |
| Ca II | 53 | 60 |
| Ca III | 20 | 50 |
| Ca IV | 12 | 40 |
| Ca V | 21 | 35 |
| Ca VI | 16 | 30 |
| Fe III | 65 | 607 |
| Fe IV | 21 | 280 |
| Fe V | 19 | 182 |
| Fe VI | 10 | 80 |

**Note.** $N_S$ and $N_F$ represent the super levels and full atomic levels used for any atomic species.

From the overall ionization structure of the atmosphere (Figure 3), we find population equilibrium zones between any two ionization states of each element. At the equilibrium zones, the electron opacity is equally contributed by both the ions. The inner wind is driven by the line opacity from the higher ionization states of neon, silicon, and iron (see Figures 3(d)–(f)). At the optical photosphere boundary ($\tau_{ross} = 0.66$), we find peaks for the higher ionization states and troughs for the lower ionization states of the various elements, which indicate a sudden change in the population fractions for these atomic species, due to strong ionization of the abundant elements by the UV radiation pressure (due to Fe opacity).

The model (see Figure 1) produced most of the line strengths (~90%) of emission lines present in the observed spectra. The under and overprediction of line strengths of some of the emission lines can be attributed to insufficient opacity data (as also suggested by Rosslowe 2016; Aadland et al. 2022) or due to the size of the chosen atomic model. We summarize the physical and chemical properties of the best-fitting stellar model atmosphere in Table 7.

### 4.1. Physical Characteristics of the Stellar Wind

To constrain the physical parameters, we compare the models based on the line strength ratios: C II/C III, C III/C IV, and He I/He II. We estimate the physical parameters considering the diagnostic emission lines: C IV $\lambda5471$, $\lambda\lambda5802$–12, C III $\lambda5696$, C II $\lambda6574$, 7231, and He II $\lambda5411$ in the optical range and He I 2.058, C IV 2.070–2.084 $\mu$m, C III+He I 2.112–2.137 $\mu$m, He II 2.165 $\mu$m, and He II 2.189 $\mu$m in the NIR range. We do not consider C III 2.325 $\mu$m, as the CMFGEN models are known (Rosslowe 2016) to be affected due to lack of sufficient atomic data.

The luminosity ($L_* = 10^{5.89} L_\odot$) of the star is estimated from both the continuum and the relative strengths of the dominant optical emission lines: C II $\lambda6574$, 7231, C III $\lambda5696$, C IV $\lambda\lambda5802$–12, and He I $\lambda5876$, 6574. The fitting errors in the luminosity are estimated from the line strengths ratios of C II $\lambda7231$/C III $\lambda5696$. The uncertainty limits are drawn as the model line ratio differs from the observed. The upper limit of luminosity is estimated at $L_* = 10^{6.05} L_\odot$ (i.e., at +0.16 dex), while the lower limit is drawn at $L_* = 10^{5.75} L_\odot$ (i.e., at −0.14 dex).

Based upon the ionization structure of the stellar wind (see Figure 3), the inner core radius of the atmosphere is determined from the ratio of C III/C IV (Hillier 1989), whereas, the outer photosphere radius is determined from the C II/C III line ratio, which is known to be more significantly influenced by variation of the model temperature as discussed in Williams et al. (2015).

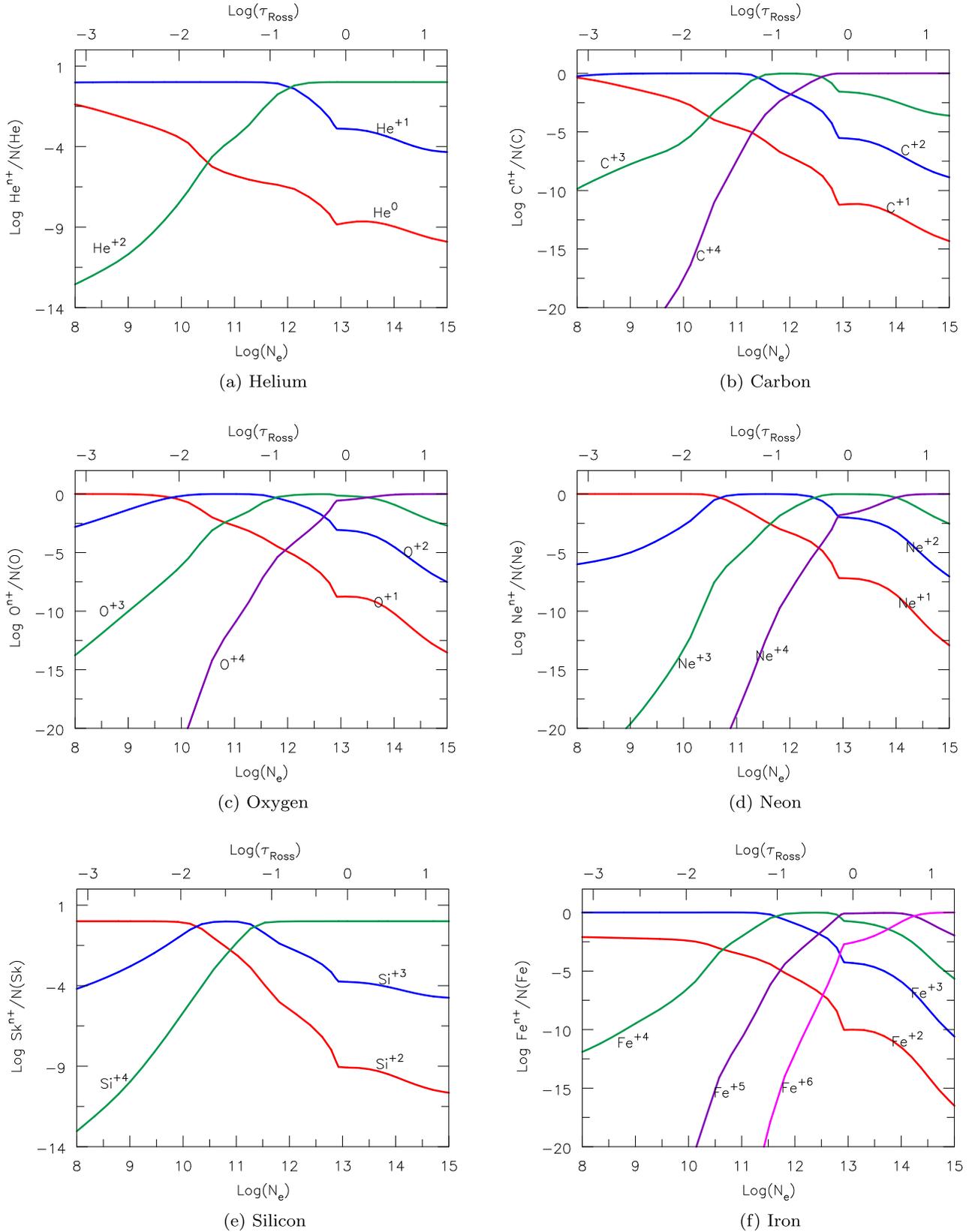

**Figure 3.** Wind ionization stratification for the considered atomic species (excluding the least abundant) in the stellar atmosphere (Section 4).

From the estimated stellar radii at different optical depths (listed in Table 7), it is found that the optical photosphere is located close to the stellar core ($R_{\tau=2/3}/R_* \simeq 1.2$). The model uncertainties in the radius are estimated from the line ratios:

**Table 7**
Spectroscopic Model Fitting Parameters

| Parameters | Values | |
|---|---|---|
| $\log L_*(L_\odot)$ | $5.89^{+0.16}_{-0.14}$ | |
| $T_*$ (K) | 44,330 | |
| $R_*(R_\odot)$ | $15.09^{+2}_{-2}$ | |
| $T_{2/3}$ (K) | 40,630 | |
| $R_{2/3}(R_\odot)$ | 17.96 | |
| $\log \dot{M}(M_\odot\ \mathrm{yr}^{-1})$ | $-4.408^{+0.042}_{-0.046}$ | |
| f | 0.1 | |
| $\log \dot{M}/\sqrt{f}(M_\odot\ \mathrm{yr}^{-1})$ | $-3.908$ | |
| $v_{\mathrm{core}}(\mathrm{km\ s}^{-1})$ | 0.4 | |
| $v_{\mathrm{phot}}(\mathrm{km\ s}^{-1})$ | 100 | |
| $v_{\mathrm{dop}}(\mathrm{km\ s}^{-1})$ | 50 | |
| $v_{\infty,1}(\mathrm{km\ s}^{-1})$ | 1000 | |
| $v_{\infty,2}(\mathrm{km\ s}^{-1})$ | 1300 | |
| $\beta_1$ | 1 | |
| $\beta_2$ | 50 | |
| $\eta$ | 2.981 | |
| $M(M_\odot)$ | 24 | |
| Atomic elements | Relative fraction | Mass fraction |
| H | 0.00 | 0.00 |
| He | 1.00 | 0.542 |
| C | $0.25^{+0.15}_{-0.24}$ | $0.406^{+0.111}_{-0.379}$ |
| N | 0.00 | 0.00 |
| O | 0.02 | 0.043 |
| Ne | $2.2e^{-3}$ | $6.02e^{-3}$ |
| Si | $2.26e^{-4}$ | $8.62e^{-4}$ |
| S | $8.75e^{-5}$ | $3.82e^{-4}$ |
| Ar | $1.92e^{-5}$ | $1.04e^{-4}$ |
| Ca | $1.12e^{-5}$ | $6.15e^{-5}$ |
| Fe | $2.18e^{-4}$ | $1.66e^{-3}$ |

C II $\lambda 7231$/C III $\lambda 5696$ and He I 2.059/He II 2.189, as these are found to be highly susceptible to the choice of the stellar radius.

The wind density and ionization structure (Rosslowe 2016) are influenced by both the mass outflow rate ($\dot{M} = 10^{-4.408}\ M_\odot\ \mathrm{yr}^{-1}$) and temperature ($T_* = 44{,}330$ K). Therefore, we derive these parameters simultaneously. The errors associated with the mass-loss rate are determined from the fitting of C II $\lambda 6578$/C III $\lambda 5696$ as it is strongly affected by the same.

The clumping factor $f_\infty$ as 0.1 and $v_{\mathrm{clump}}$ as 100 km s$^{-1}$ are chosen based on the fitting of electron scattering wings of the lower ionization state of the elements. Clumping affects the density and mass distribution in the atmosphere; hence, the estimation of a reduced (due to clumping) mass-loss rate ($\dot{M}/\sqrt{f} = 10^{-3.908}\ M_\odot\ \mathrm{yr}^{-1}$) is necessary (Crowther et al. 2006).

We adopt a two-component velocity law to address the varying line profiles from the lower and higher excited ions. The wind terminal velocity ($v_{\infty,1} = 1000$ km s$^{-1}$; $v_{\infty,2} = 1300$ km s$^{-1}$), microturbulent velocity ($v_{\mathrm{dop}} = 50$ km s$^{-1}$), and acceleration component ($\beta$) are estimated by fitting the line widths of both the lower and higher ionization states of the atomic species emitting in the optical and NIR wave bands such as that of C III and C II, and He II and He I simultaneously. The expanding stellar wind follows a velocity profile, as shown in Figure 2(a). It depicts that the wind acceleration is higher in the inner atmospheric layers, while it becomes almost zero from $100\ R_*$. The wind velocity (using Equation (1)) at the photosphere boundary is found to be 87 km s$^{-1}$. However, the emission lines are produced from an extended region of the stellar wind and are mostly broadened by the outer stellar winds.

The maximum momentum given to the wind by an ensemble of lines, assuming single scattering, is given by $L_*/c$. The wind efficiency parameter, $(\dot{M} v_\infty)/(L_* c)$, can be used to indicate the importance of multiple scattering. The estimated value shows that each photon undergoes scattering about three times before it escapes. Such a high value ($>1$) explains the reason behind the high mass-loss rate, as they enhance the momentum of the stellar wind. Due to multiple scattering phenomena, a significant amount of the radiative energy is reflected toward the photosphere of the atmosphere (also known as *wind blanketing*, Gräfener et al. 2002), which affects the temperature structure of the wind. The blanketing effect is treated using the super-level atomic transitions of the important elements.

As the hydrostatic core of the star is embedded much deeper in the optically thick region of the atmosphere, the stellar mass of the object cannot be properly estimated using the modeling code. Therefore, we follow Sander et al. (2012) and derive the mass (24 $M_\odot$) of the WC object from the mass–luminosity relationship (Langer 1989):

$$\log \frac{L_{\mathrm{WC}}}{L_\odot} = 2.971463 + 2.771634 \log \frac{M_{\mathrm{WC}}}{M_\odot}$$
$$- 0.487209 \left(\log \frac{M_{\mathrm{WC}}}{M_\odot}\right)^2$$
$$- \left[0.487870 + 0.434909 \log\left(\frac{M_{\mathrm{WC}}}{M_\odot}\right)\right.$$
$$\left. + 0.093793 \left(\log \frac{M_{\mathrm{WC}}}{M_\odot}\right)^2\right] \cdot Y, \quad (3)$$

where $L_{\mathrm{WC}}$ and $M_{\mathrm{WC}}$ represent the luminosity and the mass of the object, while $Y$ represents the helium mass fraction.

### 4.2. Chemical Composition

We estimate the surface chemical composition by fitting the set of emission lines present in the observed spectrum.

As C IV $\lambda 5471$ and He II $\lambda 5411$ are, in general, weak in WCL-type stars, it is difficult to derive the surface carbon abundance accurately using such diagnostic lines. Nevertheless, we estimate the best possible values for carbon abundance upon comparing the model flux values of most of the prominent emission lines of C III $\lambda 5696$, 6727–73, 8198, 8348, 8500, and 8664. We use the C III $\lambda 5696$/He I $\lambda 5876$ line ratio to establish the uncertainty limits to the relative carbon abundance. We establish the lower limit when the model and observed lines are reversed or equal in strength. The upper limit is established when the model line ratio deviates from that of the observed (see Figure 4).

The determination of oxygen abundance is much more challenging than that of carbon due to the lack of stronger emission lines. Due to blending with the adjacent lines of carbon, there are few emission lines of oxygen available to analyze in the optical spectrum, such as O II $\lambda 4415$–4417 and O III $\lambda 3754$–3759, 5592. Due to poor S/N in the spectral region between 4000 and 5000 Å (see Section 2.4.1), we do not detect the former two spectral lines. Therefore, we adopt the oxygen abundance from Sander et al. (2012).

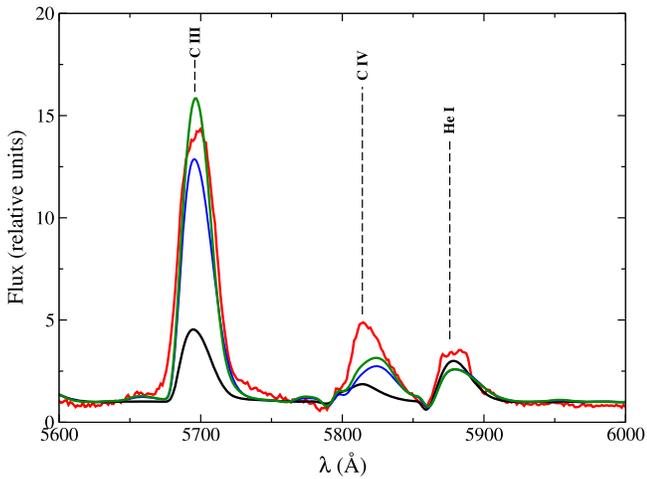

**Figure 4.** Uncertainty in the relative carbon abundance (C/He) is estimated from the fitting of the line ratio of C III λ5696/He I λ5876. The observed spectra (in red), best-fitted model (C/He = 0.25 in blue), lower uncertainty model (C/He = 0.01 in black), and the upper uncertainty model (C/He = 0.4 in green) are shown (see Table 7).

The neon abundance is adapted from an earlier study by Crowther et al. (2006) of WC9-type stars. For other heavier elements viz. silicon, sulfur, argon, calcium, and iron, we consider values from earlier studies (Hillier & Miller 1999; Smith & Houck 2005; Crowther et al. 2006; Ignace et al. 2007). We list both the relative chemical abundances and mass fractions of the atomic elements in Table 7.

### 4.3. Spectral Energy Data Fitting

To estimate the extinction parameters, we adopt the updated distance to the source as 4.33 kpc (Rate & Crowther 2020) estimated from GAIA DR2 (Bailer-Jones et al. 2018; Gaia Collaboration et al. 2018) and scale the CMFGEN model spectrum. We use the average extinction law in Gordon et al. (2023) from the Python distributed package called dust_extinction, which was comprised based on the data in Fitzpatrick et al. (2019) for dereddening the optical and from Gordon et al. (2021) for the NIR-MIR data. As the object is located in the high-extinction Galactic region, we vary the color excess parameter, $E_{B-V}$, for different $R_V$ values (within the allowed range of 2.3–5.6) and fit the dereddened observed spectrum with the model spectrum. With $E_{B-V} = 2.5$ at $R_V = 3.55$, the observed data matches well with the model data. We also use the Cardelli et al. (1989) extinction law (for $R_V = 2$–6) to check for any possible deviation in the optical-NIR extinction along the line of sight. Similarly, to check for any possible variation in the MIR extinction, we use the Chiar & Tielens (2006) extinction law for local ISM, which is mainly used to extinguish the silicate absorption features that peak at around 9.7 and 18 μm along the line of sight of most of the Galactic WC-type stars. We find no significant difference in the derived Interstellar extinction due to the choice of different average extinction laws. From the estimated extinction magnitude ($A_V$), we calculate the absolute visual-band magnitude ($M_V$) of the object (Table 8).

From the simultaneous fitting (Figure 5) of the observed spectra and photometric flux data, it is found that there is a small difference between the flux values of observed spectra and the photometric bands in the optical-NIR wavelength region. This could arise from the low S/N of the spectra in that region.

Using the derived extinction law and the estimated $E_{B-V}$, we deredden the archival photometric data set (integrated flux density) across multiple wave bands along with the observed NIR spectral data set and plot them with the model spectral energy data as shown in Figure 5. To estimate the extinction ($A_\lambda$) and absolute magnitude ($M_\lambda$) of the object across the NIR bands, we adopt the relations (see Table 8) from Stead & Hoare (2009) and Rieke & Lebofsky (1985) for estimating the extinction magnitudes for stars located in the regions other than the Galactic center. In the optical bands, the interstellar extinction is quite high ($A_V = 8.87$), which is supported by the fact that the object is embedded in a Galactic zone with high interstellar dust. However, the extinction in the NIR bands is still comparatively lower, as can be seen from the estimated parameters ($A_J = 3.04$, $A_H = 1.68$, and $A_{K_s} = 0.98$). Also, we report the NIR absolute magnitudes ($M_J = -6.55$, $M_H = -6.23$, and $M_{K_s} = -6.36$) in Table 8.

**Table 8**
Derived Photometric Quantities for the Optical-NIR Passbands of [KSF2015] 1381-19L

| Band | Apparent Magnitude | Extinction Magnitude | Absolute Magnitude | Extinction Relation |
|---|---|---|---|---|
| $V$ | 13.86 | 8.87 | −8.18 | $A_V = R_V * E_{B-V}$ |
| $J$ | 9.66 | 3.04 | −6.55 | $A_J = 3.1 * A_{K_s}$ |
| $H$ | 8.62 | 1.68 | −6.23 | $A_H = 1.71 * A_{K_s}$ |
| $K_s$ | 7.79 | 0.98 | −6.36 | $A_{K_s} = 0.11 * A_V$ |

**Note.** The algebraic relations between the extinction magnitudes for the NIR bands are taken from Rosslowe & Crowther (2015).

#### 4.3.1. Circumstellar Environment

We find an overall good fitting (with an rms flux difference of ∼0.1) among the observed and the model data in all of the wave bands (Figure 5). Extreme mass loss leads to free–free emissions in the $H$ and $K_s$ bands. However, we do not detect any excess emission in the IR bands that arise from circumstellar dust. The absence of dust signatures is attributed to its destruction by the high luminosity of the object.

### 5. Discussion

#### 5.1. Spectroscopic Attributes

Upon comparison with the earlier investigation by Kanarek et al. (2015), we note some key differences in the $K_s$-band spectrum. Kanarek et al. (2015) detected a line feature of He II at 2.217 μm, which is not seen in our observed spectrum. We observe a prominent feature of C III at 2.325 μm detected earlier (Rosslowe 2016) in WCL-type stars. Also, we see that the FWHM values for all the emission lines remain nearly unchanged, suggesting that the wind structure is invariant. Additionally, in Section 4.3.1, we do not detect any signature of circumstellar dust, which is usually seen as excess IR emissions on the total stellar continuum. To derive differences from the WC9-dusty type stars, we compare the normalized optical spectra of our object with that of WR 119 (WC9d type), which was also observed during the same night using similar configurations and reduced using the same procedure as mentioned in Section 2.1. Upon comparison (see Figure 6), we find them to have almost similar line widths and strengths except with a few key differences, such as line strengths of C III $\lambda$5696 and C IV $\lambda\lambda$5802–12, which indicates a much-extended line formation zone as well as stronger ionization in the winds of our object; however, the strength ratio between these emission lines is almost same. Also, we note that the line width of C II $\lambda$7231 is nearly the same, indicating similar wind terminal velocity. We can clearly state that our object is just as chemically evolved as WR 119 as we find the C III $\lambda$6737/He I $\lambda$6683 line ratio to be same for both objects.

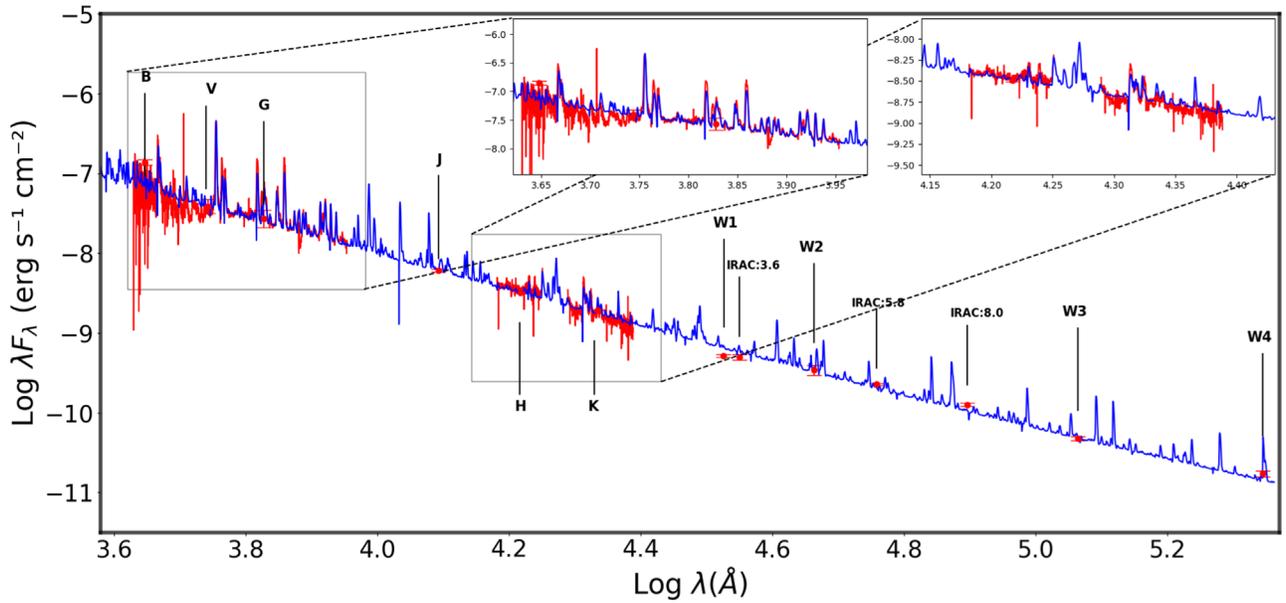

**Figure 5.** Dereddened photometric spectral energy distribution (SED; with error bars) and observed optical/NIR spectra (represented in red) are fitted with the scaled model SED (in blue) in the range between $10^{3.60}$ and $10^{5.35}$ (i.e., 4000–221000) Å. The photometric bands are also indicated with a standard symbol.

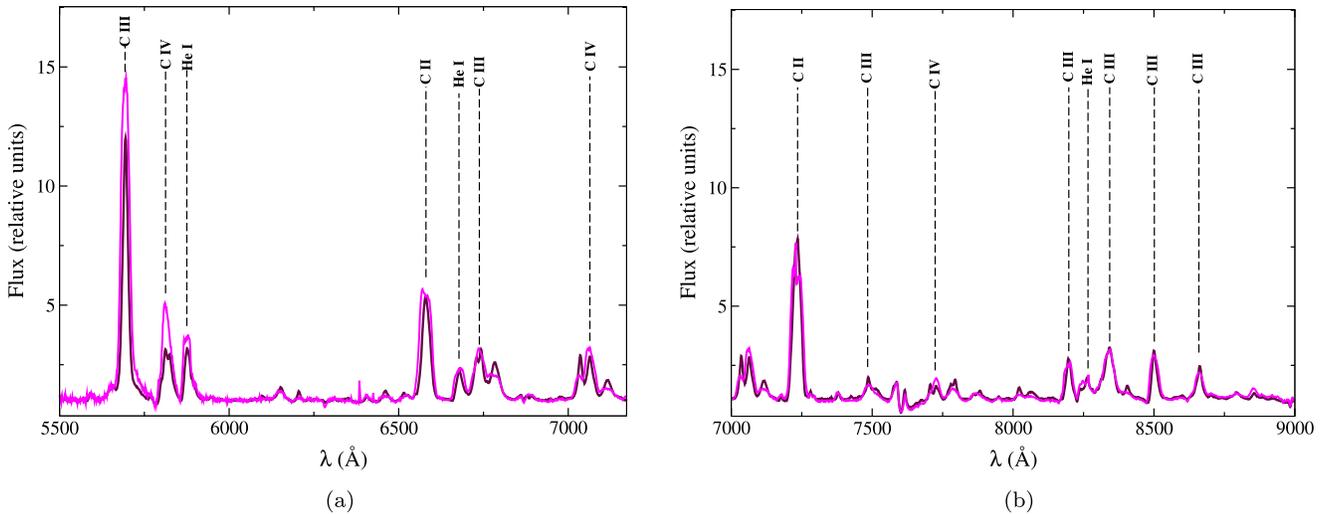

**Figure 6.** Spectral comparison between [KSF2015] 1381-19L (represented in magenta) and WR 119 (represented in maroon) across (a) 5500–7000 Å and (b) 7000–9000 Å (see Section 5.1).

A comparison of estimated stellar parameters with that of other Galactic WC9d type stars, such as WR 59, 69, 80, 103, 117, 119, and 121, is presented in Table 9. As a quick reference, we mainly consider the results from Sander et al. (2019) as they are based on the latest known distances to these objects. The object shows slightly higher luminosity than WR 59, which is suspected to be a binary (Sander et al. 2019). The temperature of the object (44,330 K) is similar to most (45,000 K) of the WC9-type objects. Also, from the estimated model parameters (Table 7), we see that there exists a temperature difference of ∼3700 K between the core ($T_*$ at $\tau \sim 20$) and the photosphere ($T_{2/3}$). Extreme mass-loss rates in WC9-type stars occur mainly due to the metallicity of the Galactic region in which the objects are located, which drives the stellar winds outward due to the Iron opacity in the inner layers of the atmosphere (Gräfener & Hamann 2005). Also, the wind terminal velocity ($v_\infty = 1300$ km s$^{-1}$) of our object is found to be slightly higher than most of the objects (in Table 9) but the same as that of WR 59 and WR 119, which we find from our spectroscopic comparison (in Figure 6).

Table 9
Comparison of [KSF2015] 1381-19L with a Few Other Galactic WC9d-type Stars

| Object | $\log L_*$ ($L_\odot$) | $T_*$ (K) | $R_*$ ($R_\odot$) | $v_\infty$ (km s$^{-1}$) | f | $\log \dot{M}$ ($M_\odot$ yr$^{-1}$) | C/He | References |
|---|---|---|---|---|---|---|---|---|
| WR 59 | 5.76 | 40,000 | 15.89 | 1300 | 0.1 | −4.48 | 0.24 | Sander et al. (2019) |
| WR 69 | 5.33 | 40,000 | 9.77 | 1090 | 0.1 | −4.87 | 0.24 | ... |
| WR 80 | 5.24 | 45,000 | 6.89 | 1600 | ... | −4.79 | ... | ... |
| WR 103 | 4.90 | 48,000 | 3.2 | 1140 | ... | −4.50 | 0.20 | Crowther et al. (2006) |
|  | 5.00 | 40,000 | 4.1 | 975 | ... | −4.50 | 0.22 | Williams et al. (2015) |
|  | 5.50 | 45,000 | 9.31 | 1190 | ... | −4.56 | 0.24 | Sander et al. (2019) |
| WR 117 | 5.36 | 56,000 | 5.12 | 2000 | ... | −4.44 | ... | ... |
| WR 119 | 4.70 | 45,000 | 3.70 | 1300 | ... | −5.13 | ... | ... |
| WR 121 | 5.16 | 45,000 | 6.35 | 1100 | ... | −4.85 | ... | ... |
| [KSF2015] 1381-19L | 5.89 | 44,330 | 15.09 | 1300 | ... | −4.41 | 0.25 | This study |

When comparing the relative elemental abundance (from Williams et al. 2015) with one of the well-studied WC9d-type stars, such as WR 103, we observe a difference of approximately 0.03 in the relative carbon abundance, which is well within the estimated model uncertainties (see Table 7). We compare our results with the PoWR spectroscopic model parameters (Sander et al. 2019), and find them to be closely aligned ($X_{He} = 0.55$, $X_C = 0.4$) taken from the PoWR WC model grid.[4] From the derived chemical abundances, the atmosphere of WCL-type stars such as [KSF2015] 1381-19L is chemically evolved, just like the WCE-type stars in the MW (Crowther et al. 2006; Sander et al. 2012). We explore the possible formation channel of isolated Galactic WC9-type objects in Section 5.2.

### 5.2. Evolutionary Phase

Evolutionary models such as Geneva,[5] MESA[6] for single stars; and Binary Population and Spectral Synthesis (BPASS)[7] models for binary systems can effectively reproduce the evolved stellar phases, such as W-R stars, luminous blue variables, red supergiants from the main-sequence progenitor stars. The single-star Geneva models consider the metallicity of the stellar environment and account for mass-loss processes that occur during the star's lifetime. The BPASS binary models evolve the primary massive star in a close interacting system with the secondary massive star via matter and momentum exchange (Eldridge et al. 2008) through RLOF. Also, BPASS binary models consider metallicity-dependent mass outflow by the stellar winds. Based on the division of the Galactic plane in terms of metallicity by Rosslowe & Crowther (2015), our object lies in the solar ($Z = 0.014$, Asplund et al. 2009) region of the Galactic disk located at an approximate Galactocentric distance ($R_G$) of 7 kpc (Kanarek et al. 2015).

For the binary evolutionary scheme, we choose the BPASS models with a fiducial IMF (slope of 1.35 and a maximum mass of 300 $M_\odot$) with all available mass ratios (0.1–0.9) and initial binary periods between 2 and 100 days, as such short-period close binaries produce W-R stars effectively (Rosslowe 2016). We retrieve the binary stellar tracks (at solar metallicity $Z = 0.014$) from the BPASS model v2.2 release (Stanway & Eldridge 2018) and use the `hoki` package (Stevance et al. 2020) for sorting the models. The BPASS binary models that undergo the WC phase are selected using the criteria from Eldridge et al. (2008), which is almost similar to the criteria in Georgy et al. (2012) used for single-star Geneva evolutionary models. The primary star (with $M > 15 M_\odot$) in a close binary evolves into a W-R star as the stellar temperature surpasses $10^{4.45}$ K and the hydrogen mass fraction ($X_H$) falls below 0.4 (Eldridge et al. 2017). The star remains in the WN phase until $X_H$ falls below 0.001. The star enters the WC phase as the total relative mass fraction of carbon and oxygen is greater than 0.03 (i.e., $X_C + X_O \geqslant 0.03$).

Almost 5452 models undergo the W-R phase and generate the WC phase, while only about 1475 models fall within the physical parameter (luminosity and temperature) space. Further, we select only those models (558) that have a period between 2 and 100 days, as such binaries are said to evolve via RLOF with chances of common envelope formation (Rosslowe 2016). The primary stars involved in the selected models are found to be 50–300 $M_\odot$. In Figure 7, we can see that the luminosity ($\log L_* = 5.89$, see Table 7) and temperature ($T_* = 44,330$ K) are reproduced by the binary stellar tracks provided the object is still undergoing the WN phase. This is because, in the interacting binaries, the WN phase has a longer lifetime (due to envelope stripping via RLOF) than the WC phase (Eldridge et al. 2008). Therefore, we do not find any suitable binary star models that can reproduce the physical and chemical characteristics of the object. We also retrieve models (see Figure 7(b)) with twice the solar metallicity ($Z = 0.04$) but find that neither of the stellar tracks could reproduce the temperature of the object in the WC phase. This is due to the lack of an envelope inflation process in all the BPASS models (Eldridge et al. 2017) of various metallicities.

We use GENEVA models (Ekström et al. 2012) to investigate the single-star evolutionary scenario. Using the published grid of stellar models (Yusof et al. 2022), we identify the possible stellar trajectory that can lead to the formation of this object and determine the corresponding progenitor mass. O-type MS stars, depending upon their initial mass, experience distinct stellar phases throughout their life cycle due to extreme mass loss and rotational mixing, ultimately culminating in core-collapse supernovae. Massive O-type stars ($M_{ini} > 25 M_\odot$) transition to the W-R stellar phase when their effective temperature ($T_{eff}$) surpasses $10^4$ K with hydrogen mass fraction ($X_H$) falling below 0.3 (Ekström et al. 2012; Georgy et al. 2012). They maintain a WN-type classification until $X_H$ declines to less than 0.05. Subsequently, when the mass fraction of carbon ($X_C$) exceeds that of nitrogen ($X_N$), the star enters the WC phase. From the SYCLIST[8] database, we generated stellar evolutionary tracks for different masses at the solar metallicity by interpolating the available models (40, 60, 85, and 120 $M_\odot$). We consider both nonrotating and rotating ($v_{rot,ini}$ = 300 km s$^{-1}$ as per Meynet & Maeder 2003) single-star evolutionary tracks and investigate the physical properties and surface chemical composition of the stellar evolutionary phase, which satisfies the WC-type criteria. Figure 8 shows a graphical representation of the object's position in the Hertzsprung–Russell (HR) diagram, accompanied by a comprehensive set of evolutionary tracks encompassing a range of masses from 50–100 $M_\odot$. In Figure 8, we show that a nonrotating main-sequence O-type star with an initial mass of 90 $M_\odot$ and a rotating star with a progenitor mass of 67 $M_\odot$ can reproduce the observational characteristics of the [KSF2015] 1381-19L. We present a comparative discussion between these closely matching evolutionary models and the stellar atmospheric model.

---

[4] https://www.astro.physik.uni-potsdam.de/~wrh/PoWR/powrgrid1.php
[5] https://www.unige.ch/sciences/astro/evolution/fr/recherche/geneva-grids-stellar-evolution-models/
[6] https://github.com/MESAHub/mesa
[7] https://bpass.auckland.ac.nz/9.html for binary systems.

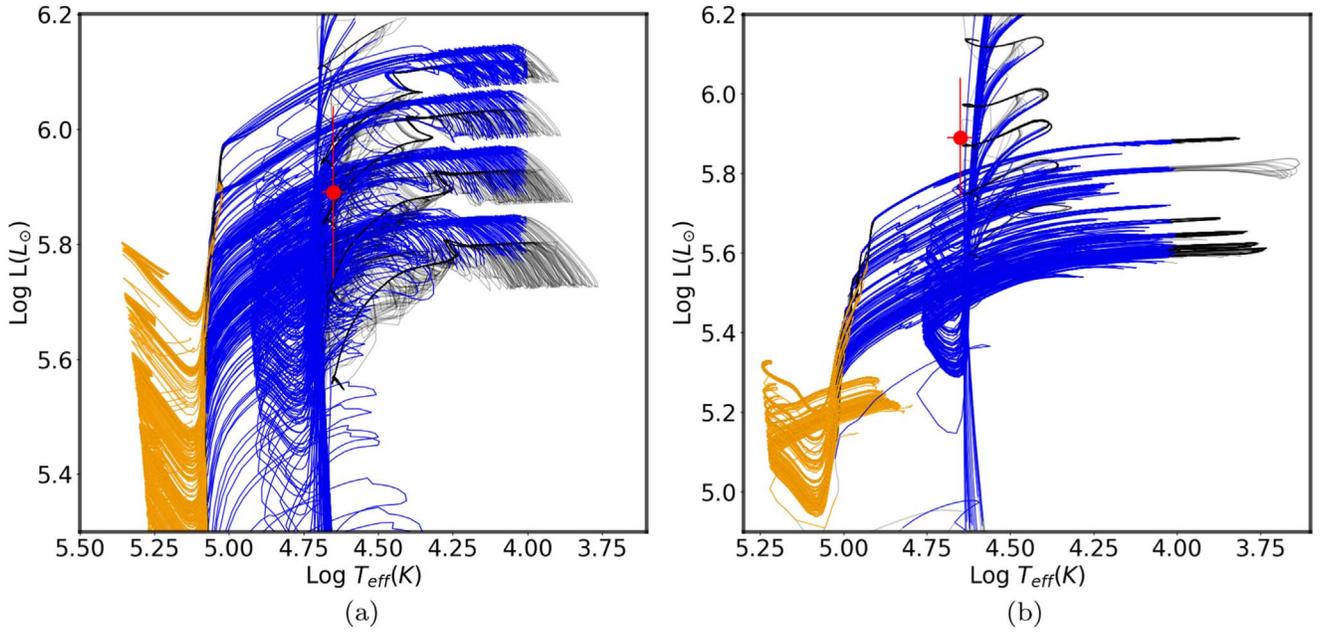

**Figure 7.** Binary stellar evolutionary models of close binaries ($P_{ini}$ = 2–100 days) at (a) solar ($Z = 0.014$) and (b) twice solar ($Z = 0.04$) metallicities. The dot (in red) symbolizes the position of the object [KSF2015] 1381-19L in the HR diagram. Different colors represent different stellar evolutionary stages (non-W-R phase in black, WN phase in blue, and WC phase in orange) in the life cycle of a massive O-type star.

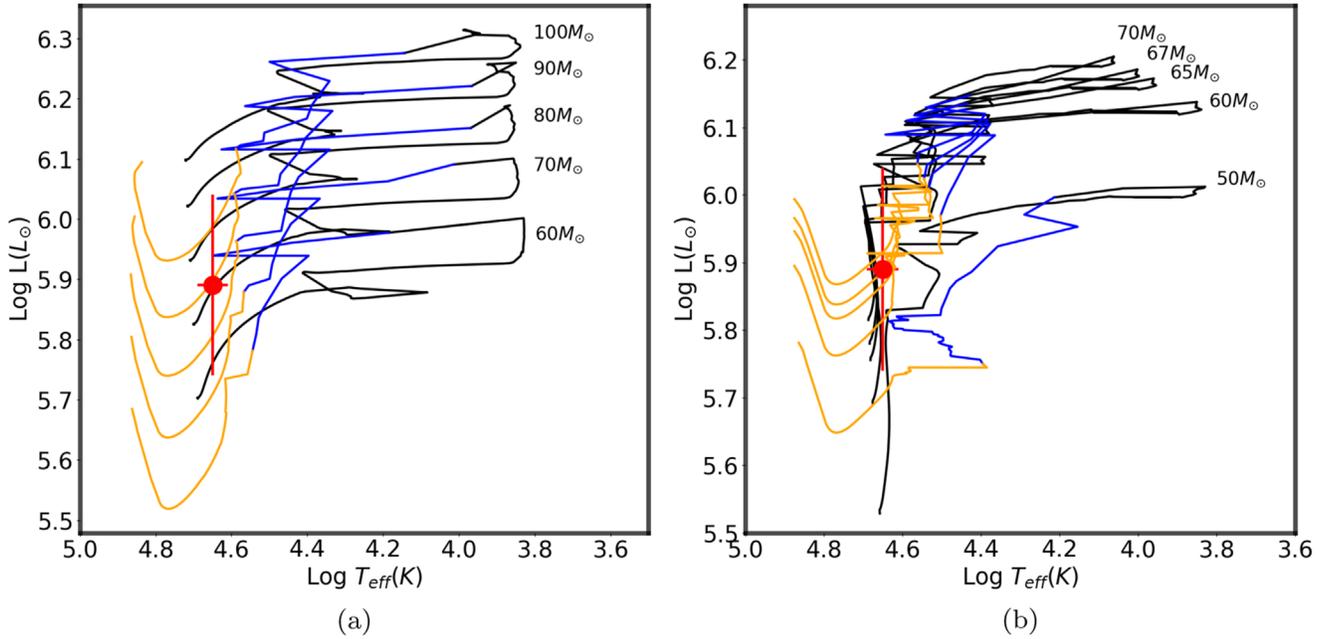

**Figure 8.** (a) Nonrotating and (b) rotating stellar evolutionary models of different masses at solar metallicity ($Z = 0.014$). We use the same color configuration as in Figure 7.

---

[8] https://www.unige.ch/sciences/astro/evolution/fr/base-de-donnees/syclist/

**Table 10**
Single-star Evolution Model Parameters for Best-fitting Nonrotating and Rotating Progenitors

| Parameter | Model | |
|---|---|---|
| | Nonrotating ($M = 90\,M_\odot$) | Rotating ($M = 67\,M_\odot$) |
| (1) | (2) | (3) |
| $\log L_*(L_\odot)$ | 5.90 | 5.89 |
| $T_*$(K) | 44258 | 44360 |
| $\log \dot{M}(M_\odot\,\mathrm{yr}^{-1})$ | −4.297 | −4.320 |
| $M(M_\odot)$ | 24.63 | 24.00 |
| $X_\mathrm{H}$ | 0 | 0 |
| $X_\mathrm{He}$ | 0.483 | 0.544 |
| $X_\mathrm{C}$ | 0.416 | 0.377 |
| $X_\mathrm{O}$ | 0.083 | 0.061 |
| $X_\mathrm{Ne}$ | 0.012 | 0.012 |

### 5.2.1. Stellar Parameters

From the model parameters (mentioned in Table 10) corresponding to the evolutionary tracks of both rotating ($67\,M_\odot$) and nonrotating ($90\,M_\odot$) models, we see that the object is undergoing core He burning and lies in the initial WC phase.

The predicted W-R masses ($\sim 24\,M_\odot$) from both rotating ($67\,M_\odot$) and nonrotating ($90\,M_\odot$) evolutionary models are close to each other and also overlap with that estimated from Equation 3; (see Section 4.1).

The mass-loss rate for the W-R phase in single-star evolutionary models is empirically related to the metallicity (Vink et al. 2001) and is therefore found to be higher ($\dot{M} = 10^{-4.32}\,M_\odot\,\mathrm{yr}^{-1}$ for the rotating $67\,M_\odot$ and $\dot{M} = 10^{-4.297}\,M_\odot\,\mathrm{yr}^{-1}$ for the nonrotating $90\,M_\odot$ models) than that predicted from the spectroscopic models ($\dot{M} = 10^{-4.408}\,M_\odot\,\mathrm{yr}^{-1}$). This is attributed to the lack of wind-clumping in the evolutionary models.

### 5.2.2. Elemental Abundances

The evolutionary models of a rotating star with an initial mass of $67\,M_\odot$ and a nonrotating star with $90\,M_\odot$ have altogether different surface mass fractions (mentioned in Table 10). Upon comparison between the elemental abundances of the evolutionary models and the spectroscopic model (Table 7), we find small differences in the most abundant species (i.e., $X_\mathrm{He}$ and $X_\mathrm{C}$), which are within the model uncertainty limits. The abundance of oxygen from the evolutionary models is larger than the spectroscopic model. Also, the $X_\mathrm{Ne}$ from the evolutionary model is higher by 0.006 than that of the spectroscopic model. These are a consequence of different nuclear reaction rates and physical conditions (such as empirical mass-loss rates) adopted in the case of the evolutionary models. Such minor differences are within the uncertainties of the spectroscopic analyses.

## 6. Summary and Conclusion

We present optical to IR characterization of a less studied WC9-type object: [KSF2015] 1381-19L located in the Galactic disk and is heavily extinguished by the dense ISM in the Galactic plane. To understand its atmospheric properties, we model the observed spectra using the CMFGEN code and derive the physical and chemical properties from the best-fitting model. From modeling, we see that the luminosity and mass-loss rate significantly affect the emission line strengths while keeping the radius fixed. The estimated stellar luminosity is close to the upper limit of the Galactic WC9-type stars, and its high mass-loss rate is influenced by the metallicity of the Galactic region. Comparing the spectroscopic properties of WR 119, we find a close alignment, indicating a chemically evolved atmosphere with a high mass-loss rate, which can support dust formation in the circumstellar environment.

In our study, we find high extinction in the optical while the estimated absolute photometric magnitudes in the NIR closely resemble those of WC9d-type stars. Excess emissions in the IR bands are attributed to the free–free emissions and not due to the circumstellar dust. However, the chances of dust formation cannot be neglected without monitoring the object in the IR bands across different epochs.

We find that the object could not have evolved in a close binary system. However, our findings from the Geneva single-star evolutionary models (at $Z = 0.014$) suggest that the object is likely to have evolved either from a rotating ($M = 67\,M_\odot$) or a nonrotating ($M = 90\,M_\odot$) O-type MS star. We predict that the object is in the initial WC phase and is undergoing core He-burning. The differences in elemental abundances are attributed to rotational mixing and nuclear reaction rates, emphasizing the importance of incorporating rotational velocity in spectroscopic modeling.

### Acknowledgments

We thank the anonymous reviewer for the valuable comments and suggestions that enhanced the manuscript significantly. We acknowledge the S. N. Bose National Center for Basic Sciences under the Department of Science and Technology (DST) of the Government of India for providing the necessary support to conduct this research work. We thank the staff of IAO, Hanle, CREST, and Hosakote, who made these observations possible. The facilities at IAO and CREST are operated by the Indian Institute of Astrophysics, Bangalore. This research has made use of the VizieR catalog access tool, CDS, Strasbourg, France. This publication makes use of data products from the 2MASS, WISE, and Spitzer (GLIMPSE survey), which is a joint project of the University of Massachusetts and the Infrared Processing and Analysis Center/California Institute of Technology, funded by the National Aeronautics and Space Administration and the National Science Foundation. This work has made use of data from the European Space Agency (ESA) mission Gaia,[9] processed by the Gaia Data Processing and Analysis Consortium (DPAC).[10] Funding for the DPAC has been provided by national institutions, in particular, the institutions participating in the Gaia Multilateral Agreement.

### ORCID iDs

Subhajit Kar https://orcid.org/0000-0001-7874-0218
Ramkrishna Das https://orcid.org/0000-0002-5440-7186
Tapas Baug https://orcid.org/0000-0003-0295-6586

---

[9] https://www.cosmos.esa.int/gaia
[10] https://www.cosmos.esa.int/web/gaia/dpac/consortium